\documentclass[aps,prl,twocolumn,superscriptaddress,floatfix,nofootinbib]{revtex4-2}
\usepackage{epsfig,amsmath,amssymb,color,comment,physics}
\usepackage[makeroom]{cancel}
\usepackage[caption=false]{subfig}
\usepackage{mathrsfs}
\usepackage[countmax]{subfloat}
\usepackage[normalem]{ulem}
\usepackage[english]{babel}
\usepackage{dsfont}
\usepackage{float}
\usepackage[bookmarks=true,colorlinks,linkcolor=black,urlcolor=NavyBlue,citecolor=RoyalBlue]{hyperref}
\usepackage[dvipsnames]{xcolor}
\usepackage{soul} 
\usepackage{color, xcolor} 

\newcommand{\doubleoverline}[1]{\bar{\bar{#1}}}
\usepackage{booktabs}
\hypersetup{
    unicode=false,     
    pdftoolbar=false,  
    pdfmenubar=true,   
    pdffitwindow=false, 
    pdfstartview={FitH},
    pdftitle={},    
    pdfauthor={Authors},     
    pdfsubject={},   
    pdfcreator={},   
    pdfproducer={}, 
    pdfnewwindow=true,
    colorlinks=true,
    linkcolor=black,
    citecolor=blue, 
    filecolor=magenta,
    urlcolor=blue
}

\newcommand{\beginsupplement}{%
    \setcounter{table}{0}
    \renewcommand{\thetable}{S\arabic{table}}%
    \setcounter{figure}{0}
    \renewcommand{\thefigure}{S\arabic{figure}}%
    \setcounter{equation}{0}
    \renewcommand{\theequation}{S\arabic{equation}}%
    \setcounter{section}{0}
    \renewcommand{\thesection}{S\arabic{section}}%
   }

\usepackage{color}

\graphicspath{{./Figures/}} 

\begin{document}
	
\title{Unconventional Distance Scaling of Casimir-Polder Force between Atomic Arrays}
	
\author{Qihang Ye} 
\thanks{These authors contributed equally}
\affiliation{School of Physics, Zhejiang Key Laboratory of Micro-nano Quantum Chips and Quantum Control, Zhejiang University, Hangzhou $310027$, China}

\author{Haosheng Fu} 
\thanks{These authors contributed equally}
\affiliation{School of Physics, Zhejiang Key Laboratory of Micro-nano Quantum Chips and Quantum Control, Zhejiang University, Hangzhou $310027$, China}

\author{Bing Miao}\email{bmiao@ucas.ac.cn}
\affiliation{Center of Materials Science and Optoelectronics Engineering, College of Materials Science and Opto-Electronic Technology, University of Chinese Academy of Sciences (UCAS), Beijing $100049$, China}
        
\author{Lei Ying}\email{leiying@zju.edu.cn}
\affiliation{School of Physics, Zhejiang Key Laboratory of Micro-nano Quantum Chips and Quantum Control, Zhejiang University, Hangzhou $310027$, China}

\begin{abstract}
Conventionally, dispersion forces mediated by quantum vacuum fluctuations are known to exhibit universal distance scalings, with retardation typically leading to a faster decay of the interaction. Here, we show that this expectation fails for intrinsically discrete systems. Using the microscopic scattering approach, we study the Casimir-Polder interaction between two atomic arrays, and uncover an unconventional distance scaling in which the force crosses over from a faster decay at short separations to a slower decay in the retarded regime. This behavior originates from the discrete lattice structure and can be consistently understood within the scattering picture. Extending our analysis to Rydberg atomic arrays, we predict an even stronger deviation from conventional scaling and propose an experimentally feasible scheme for direct measurement. Our results provide a new platform for exploring dispersion forces beyond the continuum limit.
\end{abstract}	
\maketitle

\emph{Introduction---}Quantum electrodynamics predicts that the vacuum is not empty, but instead permeated by fluctuating electromagnetic fields, which can induce dispersion forces between two neutral macroscopic objects as well as microscopic systems.
In 1948, Casimir and Polder derived the long-range dispersion interactions between neutral atoms and between an atom and a conducting surface using perturbation theory~\cite{casimir1948influence}. Subsequently, Casimir interpreted such interactions in terms of vacuum fluctuations and predicted the attractive force between two perfectly conducting plates~\cite{Casimir1948Plates}. While Casimir's original treatment was restricted to idealized conducting surface, Lifshitz formulated a general theory based on a model of fluctuating currents in realistic materials, providing a unified description of dispersion forces between macroscopic objects~\cite{lifshitz1956lifshitz,Klimchitskaya2009Casimir}. Building on this framework, dispersion interactions---exemplified by the Casimir and Casimir-Polder (CP) forces---have been extensively studied theoretically across a wide range of geometries and distance regimes~\cite{vankampen1968casimir,langbein1974theory,balian1977electromagnetic,balian1978electromagnetic}. Since the Casimir force was experimentally measured in a series of well-designed experiments~\cite{MohideenRoy1998,Bressi2002PRL,BenderCourteilleMarzokZimmermannSlama2010}, the study of dispersion forces has attracted considerable interest as a fundamental manifestation of quantum vacuum fluctuations~\cite{rodriguez2011casimir, JiangQingDong2019PRB, MiaoBing2020}.

Beyond their conceptual significance, a central feature of dispersion interactions lies in their pronounced dependence on geometry and length scale. Due to the finite speed of light, the scaling of the Casimir force changes with distance. A characteristic feature of Casimir physics is that the retardation effect modifies the distance scaling of Casimir force from $h^{-\beta}$ in the non-retarded regime to $h^{-(\beta+1)}$ in the retarded regime due to the relativistic correction~\cite{Buhmann2012,Woods2016}, where $\beta$ is the scaling exponent in the non-retarded regime. Such behavior has been confirmed for a variety of continuous geometries. For example, the interaction between a neutral atom and a thick conducting plate crosses over from the non-retarded force scaling $F \propto h^{-4}$ to the retarded force $F \propto h^{-5}$, while for macroscopic plates the force changes from $F \propto h^{-3}$ to $F \propto h^{-4}$~\cite{,milton2001casimir}. Comparable scaling crossovers have also been reported in other continuous configurations such as sphere–plate and surface geometries~\cite{MohideenRoy1998,Decca2007CasimirPressure}. Notably, these well-established results are almost exclusively based on continuous systems, while dispersion forces in intrinsically discrete systems have received much less attention.

On the other hand, with the rapid advancement of atom manipulation techniques, the construction and control of large-scale atomic arrays have become increasingly mature. Optical tweezers and optical lattice technologies now allow for the deterministic trapping and arrangement of individual atoms in arbitrary geometries~\cite{Barredo2016AtomByAtom,kumar2018prx,Hartung2024_PRA}.
The introduction of holographic and magic-wavelength tweezers has significantly improved the spatial uniformity and trapping stability of these arrays, enabling the realization of defect-free structures with high filling fractions~\cite{Liu2024_PRR,Chomaz2024_PRL}. Moreover, quite recently, a number of experiments have demonstrated highly scalable atomic arrays suitable for exploring many-body quantum physics and long-range interactions~\cite{Kaufman2024_PRXQ,Chomaz2022_SciPost}.
From this perspective, atomic arrays have emerged as a distinct class of quantum matter, characterized by programmable geometry, controllable internal states, and tunable light–matter coupling. Despite their growing importance in many-body quantum simulations~\cite{bloch2008many,browaeys2020many} and quantum-optical studies~\cite{Endres2016AtomByAtom}, the CP interactions within and between such discrete arrays remain largely unexplored.

In this paper, we study the CP interaction in systems consisting of two separated atomic arrays. Within the virtual-photon picture and the Green‘s function (GF) technique that accounts for scattering inside the arrays, we calculate the Casimir force and identify an unconventional scaling behavior. We find that, in such systems, the onset of retardation generally leads to a slower distance decay, in contrast to conventional Casimir interactions between continuous objects. 
We further propose an experimental scheme to measure this weak interaction between a Rydberg atomic array and a single Rydberg atom, which is in principle feasible with current experimental capabilities. The successful experimental verification of this force in Rydberg arrays is expected to stimulate renewed interest in Casimir interactions in discrete systems.

\begin{figure}[ht]
    \centering
    \includegraphics[width = 0.9\linewidth]{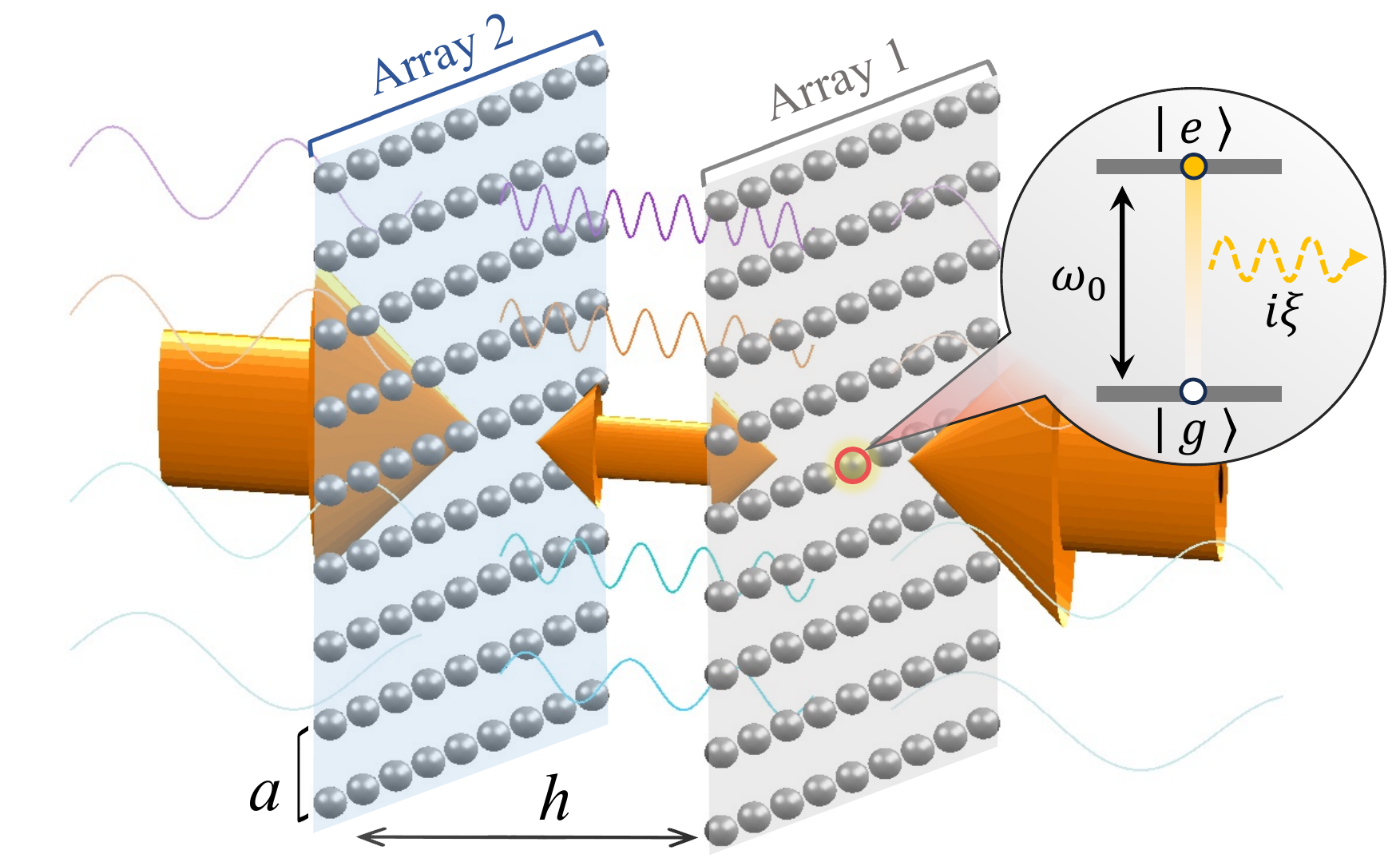}
    \caption{Illustration of the Casimir-Polder force between two 2D atomic arrays.
    The inset illustrates a quantum fluctuation that excites a two-level atom marked by the red circle in Array 1 and emits a virtual photon (dashed wavy arrow), which can be scattered by atoms in the arrays. Solid colored wavy lines indicate the virtual photons scattering between atomic arrays.}\label{fig:1}
\end{figure}
     
\emph{System and Method---}We consider two parallel two-dimensional (2D) atomic arrays in vacuum, separated by a distance $h$, as illustrated in Fig.~\ref{fig:1}. Each array consists of identical atomic two-level systems (TLSs) arranged in a square lattice with lattice constant  $a$. 
Each atom is treated as an isotropic quantum electric dipole, whose 
polarizability is~\cite{lambropoulos2007fundamentals}
$\alpha(\omega) = 2 d_0^2 \omega_0/[\hbar(\omega_0^2-\omega^2 - i\gamma\omega)]$.
Here, $\omega_0$, $\gamma$, and $d_0$ denote the transition frequency, the  spontaneous decay rate, and the dipole moment magnitude, respectively.  

In such a system, the force arises from the modification of the electromagnetic mode spectrum induced by two atomic arrays. The total Hamiltonian reads
$\hat{H} = \hat{H}_{\mathrm{A}} + \hat{H}_{\mathrm{F}} + \hat{H}_{\mathrm{AF}}$, where
$\hat{H}_{\mathrm{A}} = \sum_{n}\hbar\omega_0 \hat{\sigma}^{+}_n \hat{\sigma}^{-}_n$
describes the atomic TLSs, and
$\hat{H}_{\mathrm{F}} = \int d^3r \int_{0}^{\infty} d\omega\, \hbar\omega\,
\hat{f}^{\dagger}(\mathbf{r},\omega)\hat{f}(\mathbf{r},\omega)$
is the Hamiltonian of the quantized electromagnetic field.
Here, $\hat{\sigma}^{\pm}_n$ are the raising (lowering) operators of the $n$th TLS, while
$\hat{f}(\mathbf{r},\omega)$ and $\hat{f}^{\dagger}(\mathbf{r},\omega)$ denote the bosonic annihilation and creation operators of the elementary excitation of the combined field--medium system. The atom--field interaction Hamiltonian is given by~\cite{PhysRevA.55.1485,Buhmann2012,PhysRevLett.123.173901}
\begin{equation}
\hat{H}_{\mathrm{AF}} = \sum_{n}\Big[
-\hat{\mathbf{d}}_n\cdot\hat{\mathbf{E}}^{\parallel}(\mathbf{r}_n)
- \frac{e}{m}\hat{\mathbf{P}}_n\cdot\hat{\mathbf{A}}(\mathbf{r}_n)
+ \frac{e^2}{2m}\hat{\mathbf{A}}^2(\mathbf{r}_n)
\Big],
\end{equation}
where the dipole operator of the $n$th atom is
$\hat{\mathbf{d}}_n \equiv \mathbf{d}_0\hat{\sigma}^{+}_n + \mathbf{d}_0^{*}\hat{\sigma}^{-}_n$,
with $\mathbf{d}_0$ being the transition dipole moment. Within the macroscopic QED formalism, the vector potential is expressed as
$\hat{\mathbf{A}}(\mathbf{r}) = \int_0^\infty d\omega \int d^3r'
\sqrt{\hbar/(\pi\varepsilon_0\omega)}\doubleoverline{\mathrm{G}}^{\mathrm{(s)}}(\mathbf{r},\mathbf{r}',\omega)
\cdot \hat{\mathbf{f}}(\mathbf{r}',\omega) + \mathrm{H.c.}$, where $\doubleoverline{\mathrm{G}}^{\mathbf{(s)}}(\mathbf{r},\mathbf{r}',\omega)$ is the classical electromagnetic Green's tensor of the full system, incorporating the geometry and scattering properties of the atomic arrays. The transverse electric field follows from $\hat{\mathbf{E}}^{\parallel}(\mathbf{r}) = -\partial_t \hat{\mathbf{A}}(\mathbf{r})$, and $\hat{\mathbf{P}}_n$ denotes the canonical center-of-mass momentum operator of the $n$th atom.

Since the coupling between the TLSs and the electromagnetic environment is in the weak regime, higher-order processes involving two or more photons can be safely neglected. 
Consequently, the interaction can be accurately described within the single-photon approximation. By treating the atom–field coupling as a perturbation and taking the unperturbed state as $|0\rangle = |g\rangle|\{0\}\rangle$ with the intermediate state $|e\rangle|\{1\}\rangle$, equivalently we can focus on one atom at position $\mathbf{r}_0$ interacting with the effective field generated by all the other atoms. Then, the CP potential at position $\mathbf{r}_0$ can be written as an integral over the imaginary axis of frequency (see detailed derivations in supplemental material (SM)):
\begin{equation}
    U_{\mathrm{CP}}(\mathbf{r}_0)
    = -\frac{\hbar}{2\pi} \int_{0}^{\infty} d\xi 
    \sum_{j = x, y, z} \frac{\mathrm{d}p_{j}(\mathbf{r}_0)}{p_{0,j}},
    \label{eq:UCP}
\end{equation}
where polarization $\mathbf{p}=[p_x,p_y,p_z]^T$ is given by
\begin{equation}
\mathbf{p}(\mathbf{r}_n)
= \mathbf{p}_0(\mathbf{r}_n)
  - \sum_{m \neq n}
  \frac{4 \pi^2}{\epsilon_0 \lambda^2}
  \doubleoverline{\alpha}(i\xi)\,
  \doubleoverline{\mathrm{G}}^{\mathrm{(0)}}(\mathbf{r}_n, \mathbf{r}_m, i\xi)\,
  \mathbf{p}(\mathbf{r}_m).
  \label{eq:scattering}
\end{equation}
Here, $\doubleoverline{\mathrm{G}}^{\mathrm{(0)}}$ denotes the free-space Green's tensor~\cite{PhysRevLett.119.023603}.
The vector $\mathbf{p}(\mathbf{r}_n)$ represents the final polarization at position $\mathbf{r}_n$, while
$\mathbf{p}_{0}(\mathbf{r}_n)$ is the initial polarization, which is nonzero only for the atom located at $\mathbf{r}_0$, and $\mathrm{d}\mathbf{p}(\mathbf{r}_0)$ is the induced polarization. We denote the vacuum permittivity by $\epsilon_0$, the imaginary frequency of the virtual photon by $i\xi$, and use $\lambda = 2\pi c/\xi$ as the modulus of the virtual photon wavelength. For infinitely extended atomic layers, it is convenient to Fourier transform both the Green‘s tensor and the polarization. Combined with the Weyl decomposition~\cite{chew1999wave,PhysRevA.96.063801}, this procedure allows the problem to be solved in a semi-analytical manner, substantially improving numerical efficiency and accuracy (see SM for details).

In the following, we consider $^{87}\mathrm{Rb}$ atoms, which are well-established in atomic optical lattice experiments. For atoms of ground state, the dominant dipole transition occurs between the ground state $5S_{1/2}$ and the excited state $5P_{3/2}$, allowing the atoms to be modeled as TLSs with a transition wavelength $\lambda_0 = 780.2~\mathrm{nm}$, a decay rate $\gamma = 38.11~\mathrm{MHz}$, and a dipole matrix element $d_0 = 2.989~ e a_0$, where $e$ is the elementary charge and $a_0$ is the Bohr radius.
Here we only consider the D2 transition of $^{87}\mathrm{Rb}$ as the impact of D1 line and other transitions are negligible for the CP interaction. Moreover, in Eq.~(\ref{eq:UCP}), although the CP interaction formally involves an integration over all electromagnetic modes up to infinite frequency, the atomic response is limited by its highest transition frequency.
Thus, the integral is truncated at a physical cutoff far beyond all atomic resonances; for ground-state of $^{87}\mathrm{Rb}$, a cutoff on the order of $10^{18}\,\mathrm{Hz}$ is fully sufficient for converged calculations~\cite{PhysRevLett.103.123903,antezza2009fano}.
        
\begin{figure}[ht]
         \centering
         \includegraphics[width = \linewidth]{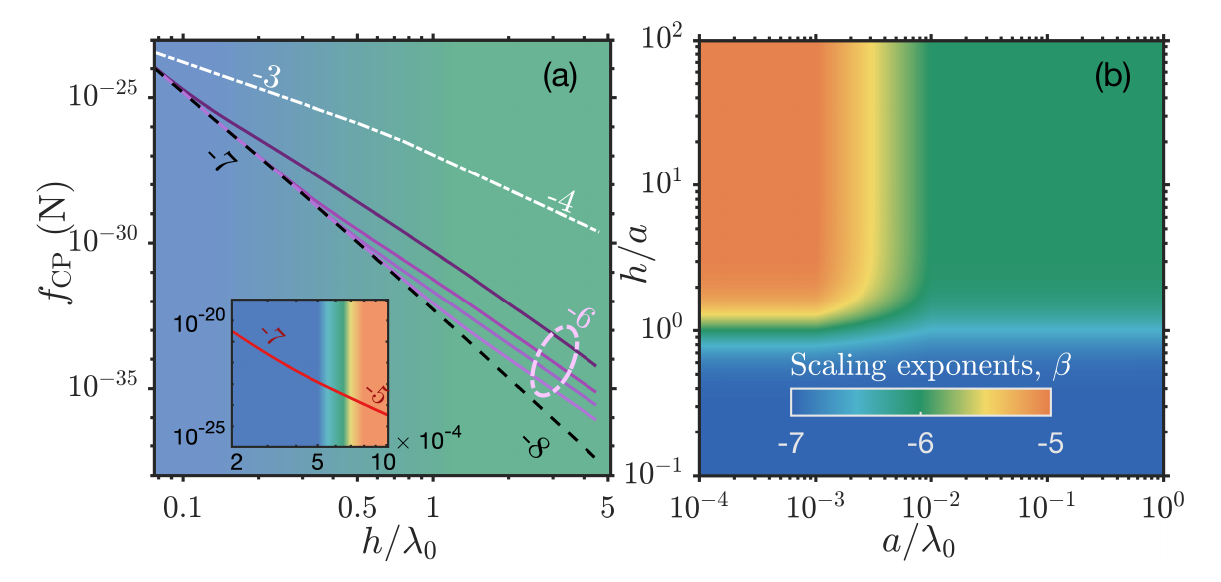}
         \caption{(a) CP or Casimir force per atom between various objects. White dash-dot line: between two conducting plate (rescaled by $10^{-4}$). Black dashed line: between two atoms. The solid lines from dark to light purple: between two ground-state Rb atomic arrays with lattice constant $a = 0.1\lambda_0$, $0.3\lambda_0$, $0.5\lambda_0$, and $0.9\lambda_0$ where $\lambda_0 = 780.2\,\mathrm{nm}$ for transition from $\mathrm{5S_{1/2}}$ to $\mathrm{5P_{3/2}}$. Numbers next to each curve indicate the power-law scaling exponent in the corresponding distance regime. Inset: the CP force between two arrays of Rydberg atoms, computed for the
        $53D_{3/2}\rightarrow 52F_{5/2}$ transition with the corresponding wavelength
        $\lambda_0 = 1.913\times10^{-2}\,\mathrm{m}$ and
        $\mu = 1.491\times10^{-26}\,\mathrm{C\cdot m}$.
        The horizontal axis and the vertical axis are both omitted since they are identical to that in Fig.~2(a).
        The lattice constant is $a = 7000\,\mathrm{nm}$.
         (b) Scaling diagram of the CP force between atomic arrays.}  
             \label{fig:2}
\end{figure}
         
\emph{Unconventional distance scaling---}Fig.~\ref{fig:2}(a) presents the distance scaling of the CP force per atom $f_{\mathrm{CP}}$ between two infinite ground-state atomic arrays with different lattice constants $a$. For comparison, the CP force between a single ground-state Rb--Rb atomic pair and the well-known Casimir force between two conducting plates are also shown as standard reference cases. 
For these reference systems, the force decays more rapidly when crossing from the nonretarded regime ($h \ll \lambda_0$) to the retarded regime ($h \gg \lambda_0$): the atom-atom interaction changes from $h^{-7}$ to $h^{-8}$, while the conducting plate exhibits a crossover from $h^{-3}$ to $h^{-4}$. To enable a direct comparison on the same plot, the metallic-plate force is rescaled by a factor of $10^{-4}$, illustrating that the CP interaction between atomic arrays is orders of magnitude weaker than that between macroscopic conductors. Remarkably, the force between atomic arrays crosses over from a rapid $h^{-7}$ decay at short distances to a slower $h^{-6}$ decay at larger separations, scaling overall with the atomic areal density.

Furthermore, the Rydberg atomic array exhibits an even more pronounced deviation from conventional scaling behavior, as shown in the inset of Fig.~\ref{fig:2}(a). In this case, the lattice constant $a$ is much smaller than $\lambda_0$ and the force crosses over from the standard Casimir–Polder scaling at short distances to a slower $h^{-5}$ decay, which occurs entirely within the nonretarded regime. Notably, the scaling exponent changes by two powers in this range, which is a distinctive feature of Rydberg atomic arrays compared with conventional Casimir systems or ordinary atomic array system.

Unlike conventional Casimir configurations characterized by a single separation length~$h$, atomic arrays introduce an additional intrinsic length scale through the lattice constant~$a$, which controls how many atoms effectively contribute to the interaction at a given separation. This extra length scale plays an important role in shaping the distinct distance-scaling behaviors of the force. Fig.~\ref{fig:2}(b) summarizes the scaling behavior of the Casimir force between 2D atomic arrays, as determined by the relative magnitudes of the lattice constant~$a$, the characteristic transition wavelength~$\lambda_0$, and the array separation~$h$. Although ground-state and Rydberg atomic arrays differ microscopically, Fig.~\ref{fig:2}(b) shows that three distinct scaling regimes can be identified based on whether the interaction is dominated by a single atom or by many atoms within the arrays, and on whether retardation effects are relevant. When the interaction is effectively dominated by a single atom, the force follows the familiar $h^{-7}$ scaling, which appears in both ground-state and Rydberg atomic arrays. The slower $h^{-6}$ scaling emerges only when the interatomic spacing is comparable to $\lambda_0$ and the array separation lies deep in the retarded regime, as realized in ground-state atomic arrays. By contrast, a distinct $h^{-5}$ scaling arises provided that the separation exceeds the lattice constant ($h\gtrsim a$), the lattice spacing is much smaller than the transition wavelength ($a\ll\lambda_0$), and the interaction remains in the nonretarded regime ($h\ll\lambda_0$), a set of conditions naturally satisfied in Rydberg atomic arrays.

We note that the $h^{-6}$ scaling of the retarded Casimir force between two 2D atomic arrays contrasts with the classic $h^{-4}$ result for perfect conductors considered by Casimir. This discrepancy arises because Casimir‘s original model assumes that the plates perfectly reflect virtual photons at all frequencies, an idealization that realistic materials generally cannot satisfy. Consequently, faster decay is commonly observed due to the weakening of confinement effects: for instance, the Casimir energy between two graphene sheets transitions from $h^{-3}$ (non-retarded) to $h^{-4}$ (retarded) scaling~\cite{PhysRevB.80.245406}, corresponding to a force decaying as $h^{-5}$. In our discrete atomic arrays, the limited reflectivity is even more pronounced: the arrays can only perfectly reflect electromagnetic waves at one specific frequencies~\cite{PhysRevLett.118.113601}. Since the Casimir interaction integrates contributions over the full frequency spectrum, its scaling law is directly altered by incomplete reflection across most frequencies.

\begin{figure}[ht]
     \centering
     \includegraphics[width =1.0\linewidth]{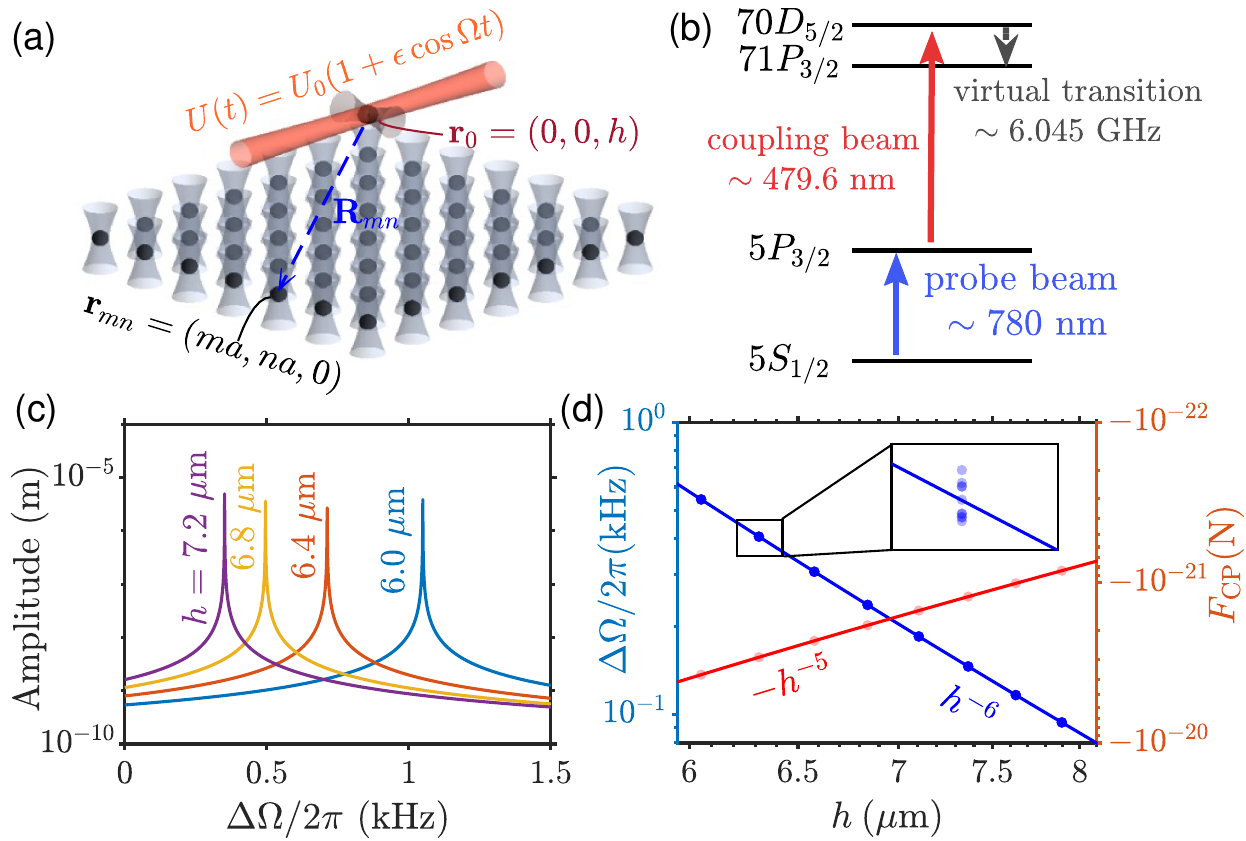}
     \caption{(a) Sketch of atom-array system.  Gray beams indicate optical trapping potentials, while red beams represent the additional modulated potential $U(t) = U_0(1+\epsilon \cos \Omega t)$. (b) Rydberg-atom excitation process, with the $70D_{5/2} \rightarrow 71P_{5/2}$ transition shown as a gray dashed line representing the virtual-photon excitation process. (c) Simulated parametric resonance of a single Rydberg atom positioned above a 2D atomic array. The plotted quantity is the atomic displacement under a periodically modulated trap; the peak amplitude indicates the resonance position. (d) Measured resonance peak shifts (left axis) and the corresponding Casimir force (right axis) as a function of atom–array distance 
    $h$. The peak shifts reflect trap curvature changes induced by the Casimir interaction, from which the force is inferred.}
     \label{fig:3}
 \end{figure}

\emph{Understanding of the scaling---}We consider a reference atom `$0$' fixed at $\mathbf{r}_0=(0,0,h)$ above a 2D atomic array $B'$, whose atoms occupy lattice sites $\mathbf{r}_{mn}=(m a,n a,0)$, as illusrated in Fig.~\ref{fig:3} (a). In this system, higher-order multiple scattering between atoms contributes negligibly, so we keep only the second-order loop. Moreover, we note that the atom–array configuration exhibits essentially the same distance scaling as the array–array system. To describe the scattering process, we define the relative displacement $\mathbf{R}_{mn}=\mathbf{r}_{mn}-\mathbf{r}_0$, which gives the separation between atom `$0$' and each atom in the lattice. The CP interaction energy can then be written as
\begin{equation}
\begin{split}
        U_{\mathrm{CP}}(h)
\approx &
-\frac{\hbar}{2\pi} \int_0^\infty d\xi\; \alpha(i\xi)^2
\sum_{m,n}\\
& \times
\mathrm{Tr}\!\left[
\doubleoverline{\mathrm{G}}^{\mathrm{(0)}}(-\mathbf{R}_{mn},i\xi)
\cdot
\doubleoverline{\mathrm{G}}^{\mathrm{(0)}}(\mathbf{R}_{mn},i\xi)
\right]\\
\approx &-\frac{\hbar}{2\pi} \int_0^\infty d\xi\; \alpha(i\xi)^2S(h;\xi),
\end{split}
\end{equation}
where $S(h;\xi) = \sum_{m,n}\exp{-2R_{mn}\xi/c }/r_{mn}^6$ represents the lattice summation over the product of two vacuum Green's tensors and  the distance scaling is completely encoded in the asymptotic behavior of 
$S(h;\xi)$.
Its form is determined by the interplay between the atom--array separation $h$ and the lattice constant $a$. In the non-retarded regime of $h\ll \lambda_0$, the GF takes a quasi-static form and decays algebraically with distance. For $h\ll a$, the summation is dominated by the nearest lattice site with $m=n=0$, yielding $S(h;\xi)\propto h^{-6}$ i.e. the CP force $F\propto h^{-7}$. In contrast, for $h \gg a$, the summation includes all lattice atoms and can be approximated by a continuum integral over the transverse plane, yielding $S(h;\xi)\propto a^{-2}h^{-4}$ and $F\propto a^{-2}h^{-5}$. Further increasing $h$ to the retarded regime of $h\gg \lambda_0$, the array geometry suppresses the effect of exponential decay in the GFs, resulting in $S(h;\xi)\propto a^{-2}h^{-5}$ and $F\propto a^{-2}h^{-6}$. Upon carrying out the virtual-frequency integration and derivative to obtain force, the asymptotic forms of the GF product perfectly account for the distance scaling observed in Fig.~2(b).

\emph{Experimental implementation---}We propose an experimentally feasible scheme to measure the Casimir interaction between a single Rydberg atom and a 2D array of Rydberg atoms via the Casimir-induced modification of the motional trapping frequency. 
$^{87}\mathrm{Rb}$ atoms are excited to the $70D_{5/2}$ state and confined in optical dipole traps. A 2D array of Rydberg atoms with a lattice spacing of $6000~\mathrm{nm}$ is placed at a tunable distance from the single atom, as illustrated in Fig.~\ref{fig:3}(a). The CP potential introduces a curvature correction to the trapping potential, resulting in an effective trapping frequency
\begin{equation}
\omega_{\mathrm{eff}} \approx 
\sqrt{\omega_0^2 + \frac{1}{m}\frac{\partial^2 U_{\mathrm{CP}}}{\partial h^2}}
\approx 
\omega_0\!\left(1+\frac{1}{2m\omega_0^2}\frac{\partial^2 U_{\mathrm{CP}}}{\partial h^2}\right).
\end{equation}
The corresponding frequency shift can be detected using parametric excitation, where the trapping laser intensity is weakly modulated as $U(t)=U_0[1+\epsilon\cos(\Omega t)]$. Resonant enhancement of the atomic motion occurs when $\Omega = 2\omega_{\mathrm{eff}}$, which can be inferred from atom loss or heating measurements~\cite{Harber2005}. Because this approach probes the curvature rather than the absolute position or energy shift, it is intrinsically insensitive to thermal and quantum motional fluctuations. The Casimir-induced curvature, quantified by 
$\partial^2 U_{\mathrm{CP}}/\partial h^2 \propto h^{-6}$, produces trap-frequency shifts decreasing from $\sim 10$~kHz to $\sim 0.1$~kHz as the distance increases from $h=6$ to $8~\mu\mathrm{m}$. As shown in Fig.~\ref{fig:3}(c), the jitter amplitude of atoms in the trap exhibits a resonance peak when $\Omega$ matches twice the trap frequency, $\Omega = 2\omega_{\mathrm{eff}}$. Fig.~\ref{fig:3}(d) shows a simulated measurement and the corresponding Casimir force derived from it. 
The measurement range is chosen to be larger than the radius of the Rydberg blockade. We estimate the total positional jitter $\sigma_{z,\rm tot}=(\sigma_{\rm thermal}^2+\sigma_{\rm zpf}^2)^{-1/2}$ from thermal motion $\sigma_{\rm thermal}=\sqrt{k_{\rm B}T/(m\omega_z^2)}$ and quantum zero-point fluctuations $\sigma_{\rm zpf}=\sqrt{\hbar/(2m\omega_z)}$,
where $k_{\rm B}$ is the Boltzmann constant, $T = 1\mu $K is the atomic temperature, $m$ is the atomic mass, and $\omega_z = 500\rm{kHz}$ is the axial trapping frequency. 
For the parameters used in our simulation, the resulting variation in the Casimir force due to the jitter of atoms within the array is below $1$\%, yielding an essentially straight line in the log--log plot over the designed range.
These results demonstrate that the system Rydberg atom platforms, owing to the large polarizability and long coherence times, provide a powerful and realistic route to directly measure the Casimir force of atom–array and even array-array interactions.

\emph{Discussions---}In this work, we theoretically study the Casimir-Polder interaction between 2D atomic arrays and reveal an unconventional distance scaling that departs from the traditional understanding that retardation effects generally lead to a faster decay of dispersion forces with distance between continuous materials. This behavior originates from the presence of an intrinsic microscopic length scale—the lattice spacing—which, together with the atomic transition wavelength, jointly determines the force scaling. 
Furthermore, we propose a realistic measurement scheme to directly probe this weak interaction between a Rydberg atomic array and a single Rydberg atom, and demonstrate that the predicted interaction is robust against experimentally realistic positional disorder. While the accessible distance range is partially limited by the Rydberg blockade~\cite{urban2009observation}, it remains sufficient to unambiguously verify the predicted unconventional scaling behavior in the retarded regime.

Compared with previous studies, our atomic-level scattering framework naturally resolves the discrete structure of the system, enables explicit control over many-body radiative processes, and provides a unified approach that can be straightforwardly generalized to arbitrary geometries and dimensionalities. The resulting formalism bridges microscopic quantum optics with fluctuation-induced forces and establishes a versatile platform for exploring Casimir and Casimir–Polder interactions beyond the continuum limit. We believe that this work establishes a general framework for understanding dispersion-force scaling in discrete forms of matter. More broadly, it opens new avenues for uncovering unconventional and nonequilibrium dispersion-force phenomena, enabled by engineered photonic environments~\cite{douglas2015quantum} and by correlations between atomic arrays~\cite{asenjo2019optical,bekenstein2020quantum}---such as entanglement---that go beyond simple photon-scattering mechanisms.

\emph{Acknowledgment}---We thank Prof. Kai Chang for valuable discussions and suggestions. This work was supported by the National Natural Science Foundation of China (Nos. 12375021, 12247101, 12575045), the Zhejiang Provincial Natural Science Foundation of China (No. LD25A050002), and the National Key Research and Development Program of China (No. 2022YFA1404203).

\bibliography{reference_Casimir}

\clearpage

\beginsupplement
\newpage
\pagenumbering{arabic}
\onecolumngrid
\begin{center}
{\bfseries\Large Supplemental Material for\\
\vspace{+0.2cm}
``Unconventional Distance Scaling of CP Force between Atomic Arrays''}\\
\vspace{+0.5cm}
{Qihang Ye, Haosheng Fu, Bing Miao, and Lei Ying}\\
\vspace{-1cm}
\end{center}
\vspace{1cm}

\setcounter{section}{0}
\setcounter{subsection}{0}
\renewcommand{\thesection}{\Roman{section}}
\renewcommand{\thesubsection}{\Alph{subsection}}

\section{I. Dispersion Forces and Casimir Interaction Framework}
\subsection{A. Dispersion Force}
 Dispersion forces are effective electromagnetic forces that occur between electrically neutral and unpolarized objects. The general dispersion force can be evaluated from the ground-state average of the Lorentz force: 
 \begin{equation}     \hat{\mathbf{F}}=\int\limits_{V}d^3r\left(\hat{\rho}\hat{\mathbf{E}}+\hat{\mathbf{j}}\times\hat{\mathbf{B}}\right).
 \end{equation}
 For the Casimir-Polder (CP) force between two ground state atoms, an alternative approach is to consider the energy shift due to atom-field coupling.
 
 The Casimir-Polder potential, from which the CP force can be easily derived, is
	\begin{equation}
		{U}_\mathrm{CP}(\mathbf{r}_A) = \Delta E(\mathbf{r}_A),
	\end{equation}
 where $\mathbf{r}_A$ is the position of atoms. By perturbation theory,
	\begin{equation}
		\Delta E = \Delta E^{(1)} + \Delta E^{(2)} + \cdots
	\end{equation}
	with first and second order
    \begin{equation}
        \Delta E^{(1)} = \langle 0 | \hat{H}_\mathrm{AF}|0\rangle, \qquad   \Delta E^{(2)} = \sum_{I \neq 0} 
\frac{\langle0|\hat{H}_\mathrm{AF}|I\rangle \langle I|\hat{H}_\mathrm{AF}|0\rangle}{E_0 - E_I}.
    \end{equation}
    Here, $\hat{H}_\mathrm{AF}$ is the Hamiltonian of atom-field interaction.  Under Born-Oppenheimer and long-wavelength approximations, the minimal coupling for a weakly magnetic and neutral atom can be written as
    \begin{equation}
		\hat{H}_\mathrm{AF} = -\hat{\mathbf{d}}\cdot\hat{\mathbf{E}}^{||}(\mathbf{r}_0)-\sum_{\alpha \in A}\frac{q_\alpha}{m_\alpha}\hat{\mathbf{p}}_{\alpha}\cdot\hat{\mathbf{A}}(\mathbf{r}_0)+\sum_{\alpha \in A}\frac{q^2_\alpha}{2m_\alpha}\hat{\mathbf{A}}^2(\mathbf{r}_0).
    \end{equation}
    	\subsection{B. Casimir Interaction}
	For an atom in ground-state positioned at $\mathbf{r}_0$, the CP potential reads\cite{Buhmann2012}
	\begin{equation}\label{eq:S9}
		U_\mathrm{CP}(\mathbf{r}_0) = \frac{\hbar\mu_0}{2\pi}\int_{0}^{\infty}d\xi\xi^2\mathrm{Tr}\left[\doubleoverline{\alpha}(i\xi)\doubleoverline{\mathrm{G}}^{\mathrm{(s)}}( \mathbf{r}_0,\mathbf{r}_0, i\xi)\right],
        \end{equation}
    where $\mu_0$ is the permeability in vacuum and $\doubleoverline{\mathrm{G}}^{\mathrm{(s)}}$ is the scattering Green's Function (GF) from $\mathbf{r}_0$ back to $\mathbf{r}_0$ for the virtual photon of frequency $i\xi$.
	The CP Force can be obtained by
	\begin{equation}
		F_{\mathrm{CP}}(\mathbf{r}_0) = -\nabla U_\mathrm{CP}(\mathbf{r}_0).
	\end{equation}
    
    \subsection{C. Polarizability}
    The general form of polarizability is a $3\times3$ tensor, and for isotropic atom, it can be simplified as
     \begin{equation}
         \doubleoverline{\alpha}(i\xi)=\alpha(i\xi)\mathbf{I}_{3\times3}
     \end{equation}
    with~\cite{lambropoulos2007fundamentals}
    \begin{equation}
        \alpha(i\xi) = \frac{2\omega_0d_{0}^2}{\hbar}\frac{1}{\omega_0^2+\xi^2+\gamma_0\xi}.
    \end{equation}
    Here, $\mathbf{I}{3\times3}$ is the identity matrix, $\omega_0$ is the transition frequency of the atom, $d_0$ is the corresponding dipole matrix element and $\gamma_0$ is the total damping rate including the spontaneous decay $\gamma$ and the non-radiative loss $\gamma_{nr}$. Since we consider atoms in free space, where non-radiative loss is negligible, we approximate $\gamma_0\simeq\gamma$ in our calculations.

\section{II. Scattering GF Formalism}
\subsection{A. Scattering GF}
We now discuss the scattering GF appearing in Eq.~(\ref{eq:S9}). 
All atoms are assumed to be identical and fixed at their lattice positions $\mathbf{r}_n$
within the two-layer array, initially in their ground states. Vacuum fluctuations can excite the atom at $\mathbf{r}_0$, causing it to emit a virtual photon. This photon may scatter off other atoms in the array and return to the emitter, 
producing a back-action described by the scattering GF. In our approach, the atoms are treated as classical dipoles, 
so the entire process is captured by the following equation
\begin{equation}\label{eq:S12}
\mathbf{p}(\mathbf{r}_0)
= -\frac{4 \pi^2}{\epsilon_0 \lambda^2}
  \doubleoverline{\alpha}(i\xi)\,
  \doubleoverline{\mathrm{G}}^{\mathrm{(s)}}(\mathbf{r}_0, \mathbf{r}_0, i\xi)\,
  \mathbf{p}_0(\mathbf{r}_0),
\end{equation}
where $\lambda = 2\pi c/\xi$ is the characteristic wavelength of the atomic transition, 
$\mathbf{p}_0(\mathbf{r}_0)$ is the initial polarization, and $\mathbf{p}(\mathbf{r}_0)$ is the resulting polarization after self-consistently including the response to both vacuum fluctuations and the scattered fields. The tensor $\doubleoverline{\mathrm{G}}^{\mathrm{(s)}}(\mathbf{r}_0, \mathbf{r}_0, i\xi)$ is the scattering GF, which characterizes the modification of the electromagnetic field at $\mathbf{r}_0$ due to reflections and multiple scattering from the atomic arrays. To invert $\doubleoverline{\mathrm{G}}^{\mathrm{(s)}}$, we need to solve for the polarization of initial atom using the following scattering equations:
    \begin{equation}\label{eq:scattering_eq}
    \mathbf{p}(\mathbf{r}_n) = \mathbf{p}_0(\mathbf{r}_n) - \sum_{m \neq n} \frac{4 \pi^2}{\epsilon_0 \lambda^2} \doubleoverline{\alpha}(i\xi) \doubleoverline{\mathrm{G}}^{
    \mathrm{(0)}
    }(\mathbf{r}_m, \mathbf{r}_n, i\xi) \mathbf{p}(\mathbf{r}_m).
    \end{equation}
    With the help of the Eq.~(\ref{eq:S12}) and (\ref{eq:scattering_eq}), we can rewrite the CP potential into a simpler form
    \begin{equation}\label{eq:simpler_u_cp}
        U_{\mathrm{CP}}(\mathbf{r}_0) = -\frac{\hbar}{2\pi}\int_{0}^{\infty}d\xi \sum_{j=x,y,z} dp_{j}(\mathbf{r}_0) / p_{0,j}.
  \end{equation}
      
\subsection{B. Dealing with infinite Bravais lattice}

    Eq.~(\ref{eq:scattering_eq}) is sufficient for solving the problem of finite number of atoms, in this section, we introduce the method for dealing with the infinite system of which the geometry is Bravais lattice. We start with the Fourier transform of the induced polarization and GF~\cite{PhysRevLett.118.113601}
        \begin{equation}\label{eq:S15}
        \mathbf{p}_0(\mathbf{k}_{||}) = \sum_{n}e^{-i\mathbf{k}_{||}\cdot\mathbf{r}_{n}}\mathbf{p}_{0}(\mathbf{r}_n)
        \end{equation}
        and
        \begin{equation}\label{eq:S16}
           \doubleoverline{\mathrm{g}}(\mathbf{k}_{||}) = \sum_{n \neq 0}e^{-i\mathbf{k}_{||}\cdot\mathbf{r}_{n}}\doubleoverline{\mathrm{G}}^{\mathrm{(0)}}(\mathbf{0}, \mathbf{r}_n,i\xi).  
        \end{equation}
Here and in the subsequent discussion, we choose $\mathbf{r}_{0}$ to be the origin for the sake of simplicity.
Then we obtain the self-consistent equation in k-space,
        \begin{equation}
            \mathbf{p}(\mathbf{k}_{||}) = \mathbf{p}_0(\mathbf{k}_{||}) - \frac{4 \pi^2}{\epsilon_0\lambda^2} \mathbf{\alpha}(i\xi) \doubleoverline{\mathrm{g}}(\mathbf{k}_{||})  \mathbf{p}(\mathbf{k}_{||}).
        \end{equation}
        By performing an inverse Fourier transform, we obtain the polarization in real space
        \begin{equation}
            \mathbf{p}(\mathbf{r}_0) = A_0\int_{1BZ}\frac{d\mathbf{k}_{||}}{(2\pi)^2}\mathbf{p}(\mathbf{k}_{||}),    
        \end{equation}
where $A_0$ is the area of the unit cell.
        
The accuracy of the method above depends on the convergence speed of the summations in Eq.~(\ref{eq:S15}) and Eq.~(\ref{eq:S16}). The convergence speed is observed to increase with the lattice constant $a$ and this method can give satisfactory result for all $a > 0.2\lambda_0$. For smaller $a$, we introduce an analytical method based on the Weyl decomposition (See Section: Analytical method in k-space).

\subsection{C. Dealing with two or more sub-lattice}
In order to calculate the CP force between the two layers, we need to deal with a scattering problem involving an atomic array with at least two sub-lattices. In the following, we will take two layers of simple Bravais lattice as the example. First we consolidate the vector or matrix for sub-lattices into a single and extended form. The polarization becomes
    \begin{equation}
    \Tilde{\mathbf{p}}_n \equiv 
    \left(p^{(1)}_{n,x},\, p^{(1)}_{n,y},\, p^{(1)}_{n,z},\, 
          p^{(2)}_{n,x},\, p^{(2)}_{n,y},\, p^{(2)}_{n,z}\right)^{T},
    \end{equation}
    where $p^{(l)}_{n,\mu}$ ($l=1,2$; $\mu=x,y,z$) denotes the $\mu$ component of the polarization
    of the $n$th atom in the $l$th array (or sub-lattice). The vector for initial polarizations
    $\Tilde{\mathbf{p}}_{n,0}$ is defined analogously.
        The matrix of GF is extended to
        \begin{equation}\label{eq:S20}
        \Tilde{\mathrm{G}}(\mathbf{0}, \mathbf{r}_n, i\xi) \equiv \left(
        \begin{array}{cc}
           \doubleoverline{\mathrm{G}}^{\mathrm{(0)}}(\mathbf{0}, \mathbf{r}_n, i\xi)  & \doubleoverline{\mathrm{G}}^{\mathrm{(0)}}(\mathbf{0}, \mathbf{r}_n-h\hat{z},i\xi) \\
            \doubleoverline{\mathrm{G}}^{\mathrm{(0)}}(\mathbf{0}, \mathbf{r}_n+h\hat{z},i\xi) & \doubleoverline{\mathrm{G}}^{\mathrm{(0)}}(\mathbf{0}, \mathbf{r}_n, i\xi)
        \end{array}\right).
        \end{equation}
        Here, $h$ is the distance between the two layers.
        The polarizability is block-diagonal, i.e.
        \begin{equation}
            \Tilde{\alpha}(i\xi) \equiv 
            \begin{bmatrix}
                \doubleoverline{\alpha}(i\xi) & \doubleoverline{0}_{3\times3}\\
                \doubleoverline{0}_{3\times3} & \doubleoverline{\alpha}(i\xi)
            \end{bmatrix},
        \end{equation}
        where $\doubleoverline{0}_{3\times3}$ denotes the $3\times3$ zero matrix.
        By substituting the quantities with their counterparts denoted by tilde in Eq.~\ref{eq:scattering_eq}, we obtain
        \begin{equation}
            \Tilde{\mathbf{p}}(\mathbf{r}_n) = \Tilde{\mathbf{p}}_{0}(\mathbf{r}_n) - \sum_{m \neq n} \frac{4 \pi^2}{\epsilon_0\lambda^2} \Tilde{\alpha}(i\xi) \Tilde{\mathrm{G}}(\mathbf{r}_m, \mathbf{r}_n, i\xi)  \Tilde{\mathbf{p}}(\mathbf{r}_m).
        \end{equation}
        The Fourier transform can also be applied to the quantities with the tilde straightforwardly and the approach can be extended to other geometry such as honeycomb lattice without difficulties.
        \subsection{D. Analytical method in k-space}
        As mentioned earlier, direct summation in real space may encounter challenges when $a$ is very small. Here, we present an analytical method to address this problem.  
        Due to the translational invariance of the free-space dyadic GF, in the following we will replace the absolute coordinates of two points by their relative displacement and, without further emphasis, also omit the frequency dependence. Throughout the following, $k$ is defined as $k = i\xi/c$.
        
        In terms of the Weyl decomposition to rewrite the GF~\cite{chew1999wave,PhysRevA.96.063801}.
        \begin{equation}
        G_{\alpha\beta}(\mathbf{r}) = \int\frac{dp_xdp_y}{(2\pi)^2}g_{\alpha\beta}(\mathbf{p}_{x-y};z)e^{i(p_xx+p_yy)},
        \end{equation}
        where $\mathbf{p}_{x-y} \equiv p_x\hat{x}+p_y\hat{y}$. We further define $p^2 = p_x^2 + p_y^2 + p_z^2$ and obtain the decomposition function
        \begin{equation}\label{eq:S24}
        g_{\alpha\beta}(\mathbf{p}_{x-y};z) = \int \frac{dp_z}{2\pi}\frac{1}{k^2}\frac{k^2\delta_{\alpha\beta}-p_{\alpha}p_{\beta}}{k^2 - p^2}e^{ip_zz}.
        \end{equation}
        For the sake of simplicity, we omit the frequency in GF and retain only the displacement vector.\\
        The GF on the diagonal block of matrix $\Tilde{\mathbf{G}}(i\xi, 0, \mathbf{r}_n)$ becomes
        \begin{equation}\label{eq:S25}
            \begin{split}
               \sum_{n\neq0}e^{-i\mathbf{k}_{||}\cdot \mathbf{r}_n} 
                G_{\alpha\beta}(\mathbf{r}_n)
                 &= \sum_{n}e^{-i\mathbf{k}_{||}\cdot \mathbf{r}_n} G_{\alpha\beta}(\mathbf{r}_n)  -G_{\alpha\beta}(\mathbf{0})
                \\
                &=\sum_{n}e^{-i\mathbf{k}_{||}\cdot \mathbf{r}_n}\int\frac{dp_xdp_y}{(2\pi)^2}g_{\alpha\beta}(\mathbf{p}_{x-y};0)e^{i\mathbf{p}_{x-y}\cdot \mathbf{r}_n}
                -G_{\alpha\beta}(\mathbf{0})\\
                &= \frac{1}{A_0}\sum_{\mathbf{K}}g_{\alpha\beta}(\mathbf{K}+\mathbf{k}_{||};0)-G_{\alpha\beta}(\mathbf{0}),
            \end{split}
        \end{equation}
        where $\mathbf{K}$ represents the reciprocal-lattice vector and $A_0$ is the area of the periodic unit cell.\\
        In the final line above, we applied the Poisson summation formula:
        \begin{equation}
            \sum_{n}e^{i(-\mathbf{k}_{||}+\mathbf{p}_{x-y})\cdot \mathbf{r}_n} = \frac{(2\pi)^2}{A_0}\sum_{\mathbf{K}}\delta^{(2)}(\mathbf{p}_{x-y}-\mathbf{k}_{||}-\mathbf{K}),
        \end{equation}
        where $\delta^{(2)}(\mathbf{p}_{x-y}-\mathbf{k}_{||}-\mathbf{K})$ is the 2D Dirac delta function.
        For the off-diagonal block, we have
        \begin{equation}\label{eq:S27}
            \sum_{n}e^{-i\mathbf{k}_{||}\cdot \mathbf{r}_n} G_{\alpha\beta}(\mathbf{r}_n\pm h\hat{z}) = \frac{1}{A_0}\sum_{\mathbf{K}}g_{\alpha\beta}(\mathbf{K}+\mathbf{k}_{||};\pm h).
        \end{equation}
        We note that both $g_{\alpha\beta}(\mathbf{K}+\mathbf{k}_{||};0)$ and $G_{\alpha\beta}(\mathbf{0})$ exhibit divergence. This is because we do not account for the quantum fluctuations of atoms. In order to eliminate this divergence, we apply the following regularization~\cite{PhysRevLett.103.123903}
        \begin{equation}\label{eq:S28}
            G^{*}_{\alpha\beta}(\mathbf{0}) = \int d^3\mathbf{q}\:G_{\alpha\beta}(-\mathbf{q})\cdot \frac{1}{(\sqrt{2\pi}a_\mathrm{ho})^3}e^{-q^2/2a_\mathrm{ho}^2}
        \end{equation}
        and
        \begin{equation}\label{eq:S29}
            g^{*}_{\alpha\beta}(\mathbf{p}_{x-y};0) = \int \frac{dp_z}{2\pi}\frac{1}{k^2}\frac{k^2\delta_{\alpha\beta}-p_{\alpha}p_{\beta}}{k^2 - p^2+i\epsilon}e^{-a_\mathrm{ho}^2p^2/2}.
        \end{equation}
        where $a_\mathrm{ho}$ is the fluctuation of atom positions around the lattice site.
        
        With the regularization, the summation of Eq.~(\ref{eq:S25}) becomes
        \begin{equation}
           \sum_{n\neq0}e^{-i\mathbf{k}_{||}\cdot \mathbf{r}_n} G_{\alpha\beta}(\mathbf{r}_n) = \frac{1}{A_0}\sum_{\mathbf{K}}g_{\alpha\beta}^{*}(\mathbf{K}+\mathbf{k}_{||};0)-G_{\alpha\beta}^{*}(\mathbf{0}).
        \end{equation}
        Thus, for the layer containing the reference atom, we have
        \begin{equation}
            p_{0,\alpha}^{(1)}(\mathbf{k}_{||}) = \frac{1}{A_0}\left(\sum_{\mathbf{K}}g_{\alpha\beta}^{*}(\mathbf{K}+\mathbf{k}_{||};0)-G_{\alpha\beta}^{*}(\mathbf{0})\right)\frac{4 \pi^2}{\epsilon_0\lambda^2} \alpha(i\xi)p_{a,\beta}.
        \end{equation}
        For the layer not containing that atom, 
        \begin{equation}
          p_{0,\alpha}^{(2)}(\mathbf{k}_{||}) = -\frac{1}{A_0}\sum_{\mathbf{K}}g_{\alpha\beta}(\mathbf{K}+\mathbf{k}_{||};h)\frac{4 \pi^2}{\epsilon_0\lambda^2} \alpha(i\xi)p_{a,\beta}.  
        \end{equation}\\       
        Now we turn to deal with the integrals in Eqs.~(\ref{eq:S24}), (\ref{eq:S28}) and (\ref{eq:S29}). 
        The GF of free space is written as~\cite{PhysRevLett.119.023603}
        \begin{equation}\label{eq:Greensfunction}
            \begin{aligned}
            G_{\alpha\beta}(i\xi, \mathbf{r}) = -\frac{e^{ikr}}{4\pi r}&\left[ \left(1+\frac{i}{kr}-\frac{1}{(k r)^2}\right)\delta_{\alpha\beta}\right.\\
            &+\left.\left(-1-\frac{3i}{kr}+\frac{3}{(k r)^2}\right)\frac{r_\alpha r_\beta}{r^2}\right]+\frac{\delta_{\alpha\beta}\delta^{(3)}(\mathbf{r})}{3k^2}.
        \end{aligned}
        \end{equation}
        Plug it into Eq.~(\ref{eq:S28}) and finish the integral, we obtain
        \begin{equation}
            G_{\alpha\beta}^*(\mathbf{0}) = 
            \frac{k}{6\pi}\left(\frac{\mathrm{erfi}\left(\frac{k a_\mathrm{ho}}{\sqrt{2}}\right)-i}{e^{(k a_\mathrm{ho})^2/2}}-\frac{-1/2+(k a_\mathrm{ho})^2}{(\pi/2)^{1/2}(k a_\mathrm{ho})^3}\right)\delta_{\alpha\beta}, 
        \end{equation}
        where $\mathrm{erfi}$ stands for the imaginary error function.\\
        In the following, we list the result of Eq.~(\ref{eq:S24}) and (\ref{eq:S29}) with the definition $\Lambda \equiv \left(k^2-p_x^2-p_y^2\right)^{1/2}$.\\
        For $z \neq 0$, 
        \begin{equation}
        \begin{aligned}
        g_{xx}(z) &= \frac{(k^2-p_x^2)I_0(\Lambda^2; z)}{2\pi k^2},   \qquad\qquad\ \ \ 
        g_{yy}(z) = \frac{(k^2-p_y^2)I_0(\Lambda^2; z)}{2\pi k^2},\\ 
        g_{xy}(z) &= g_{yx}(z)=\frac{-p_x p_y I_0(\Lambda^2; z)}{2\pi k^2},\qquad 
        g_{xz}(z) = g_{zx}(z)=\frac{-p_x I_1(\Lambda^2; z)}{2\pi k^2},\\
        g_{yz}(z) &= g_{zy}(z)=\frac{-p_y I_1(\Lambda^2; z)}{2\pi k^2}, \qquad\quad\;
        g_{zz}(z) = \frac{k^2 I_0(\Lambda^2; z)- I_2(\Lambda^2; z)}{2\pi k^2},
        \end{aligned}
        \end{equation}
        where
        \begin{equation}
            \begin{aligned}
                I_0(b;z) &= \int_{-\infty}^{+\infty}dq \frac{e^{iqz}}{b-q^2}
        =\mathrm{Re}\left[-\frac{\pi}{\sqrt{-b}}\ e^{-\sqrt{-b}|z|}\right],\\
                I_1(b;z) &= \int_{-\infty}^{+\infty}dq \frac{qe^{iqz}}{b-q^2}= i\:\mathrm{Im}\left[-i\pi\:\mathrm{sgn}(z)\:e^{-\sqrt{-b}|z|}\right],\\
                I_2(b;z) &=  \int_{-\infty}^{+\infty}dq \frac{q^2e^{iqz}}{b-q^2}=\mathrm{Re}\left[\pi\sqrt{-b}\ e^{-\sqrt{-b}|z|}\right].\\
            \end{aligned}
        \end{equation}
        For $z = 0$, we have
        \begin{equation}
        \begin{aligned}
        &g^*_{xx}(0) = \frac{(k^2-p_x^2)I^*_0(\Lambda^2; 0)}{2\pi k^2}e^{a_\mathrm{ho}^2(\Lambda^2-k^2)/2},  \qquad\qquad \;\;
        g^*_{yy}(0) = \frac{(k^2-p_y^2)I^*_0(\Lambda^2; 0)}{2\pi k^2}e^{a_\mathrm{ho}^2(\Lambda^2-k^2)/2},\\ 
        &g^*_{xy}(0) = g_{yx}(z)=\frac{-p_x p_y I^*_0(\Lambda^2; 0)}{2\pi k^2}e^{a_\mathrm{ho}^2(\Lambda^2-k^2)/2}, \quad\quad
        g^*_{xz}(0) = g_{zx}(z)=\frac{-p_x I^*_1(\Lambda^2; 0)}{2\pi k^2}e^{a_\mathrm{ho}^2(\Lambda^2-k^2)/2},\\
        &g^*_{yz}(0) = g_{zy}(z)=\frac{-p_y I^*_1(\Lambda^2; 0)}{2\pi k^2}e^{a_\mathrm{ho}^2(\Lambda^2-k^2)/2},\qquad\quad\;
        g^*_{zz}(0) = \frac{k^2 I^*_0(\Lambda^2; 0)- I^*_2(\Lambda^2; 0)}{2\pi k^2}e^{a_\mathrm{ho}^2(\Lambda^2-k^2)/2},
        \end{aligned}
        \end{equation}
        where
        \begin{equation}
            \begin{aligned}
               & I^*_0(b;0) = \int_{-\infty}^{+\infty}dq\frac{e^{-a_\mathrm{ho}^2q^2/2}}{b-q^2} 
                = \mathrm{Re}\left\{\frac{-\pi \left[1-\mathrm{erf}\left(\frac{\sqrt{-b}a_\mathrm{ho}}{\sqrt{2}}\right)\right]}{\sqrt{-b}e^{ba_\mathrm{ho}^2/2}}\right\},
                \ \ \ \ \ \ \ \ \ \ \ \ \ \ \ \ \ I^*_1(b;0) = \int_{-\infty}^{+\infty}dq\frac{qe^{-a_\mathrm{ho}^2q^2/2}}{b-q^2} = 0,\\
                &I^*_2(b;0) = \int_{-\infty}^{+\infty}dq\frac{q^2e^{-a_\mathrm{ho}^2q^2/2}}{b-q^2} = \mathrm{Re}\left\{-\frac{\sqrt{2\pi}}{a_\mathrm{ho}}+\frac{\pi \sqrt{-b}\left[1-\mathrm{erf}\left(\frac{\sqrt{-b}a_\mathrm{ho}}{\sqrt{2}}\right)\right]}{e^{ba_\mathrm{ho}^2/2}}\right\},
            \end{aligned}
        \end{equation}
        with $\mathrm{erf}(x)$ being the error function in complex plane.
        
        The expressions above are valid for both real and imaginary values of $k$, allowing us apply them to evaluate the CP potential in both non-resonance and resonance cases.
                     
\section{III. Scaling of CP Forces from a Scattering Perspective}

In this section we analyze the distance scaling of the CP force using a scattering formulation, and clarify the origin of the different power laws for atom--atom and atom--array configurations.

\subsection*{A. Atom--Atom interaction}

First, we consider an isotropic atom '$0$' located next to another atom $B$, with the two atoms separated by a distance $h$ along the $z$ axis. Without loss of generality, we choose the coordinate origin at atom $B$ and place atom '$0$' at a height $h$ above it, such that
\begin{equation}
\mathbf{r}_0=(0,0,h), \qquad \mathbf{r}_B=(0,0,0).
\end{equation}
Within the scattering approach to CP interactions, the induced dipole moments of the two atoms satisfy the coupled equations
\begin{align}
\mathbf{p}(\mathbf{r}_0)
&=
\mathbf{p}_0(\mathbf{r}_0)
-
\frac{4\pi^2}{\epsilon_0\lambda^2}
\,\alpha(i\xi)\,
\doubleoverline{\mathrm{G}}^{\mathrm{(0)}}(\mathbf{r}_0 - \mathbf{r}_B,i\xi)\,
\mathbf{p}(\mathbf{r}_B),
\label{eq:pA_scattering}
\\
\mathbf{p}(\mathbf{r}_B)
&=
\mathbf{p}_0(\mathbf{r}_B)
-
\frac{4\pi^2}{\epsilon_0\lambda^2}
\,\alpha(i\xi)\,
\doubleoverline{\mathrm{G}}^{\mathrm{(0)}}(\mathbf{r}_B - \mathbf{r}_0,i\xi)\,
\mathbf{p}(\mathbf{r}_0).
\label{eq:pB_scattering}
\end{align}
Noting that $\mathbf{p}_0(\mathbf{r}_B) = \mathbf{0}$, we can substitute Eq.~\eqref{eq:pA_scattering} into \eqref{eq:pB_scattering} to solve $\mathbf{p}(\mathbf{r}_0)$. With the help of Eq.~(\ref{eq:simpler_u_cp}), we have
\begin{equation}
U_{\mathrm{CP}}(h)
=
-\frac{\hbar}{2\pi}
\int_0^\infty d\xi\;
\alpha(i\xi)^2\,
\mathrm{Tr}
\!\left[
\doubleoverline{\mathrm{G}}^{\mathrm{(0)}}(\mathbf{r}_0- \mathbf{r}_B,i\xi)
\cdot
\doubleoverline{\mathrm{G}}^{\mathrm{(0)}}(\mathbf{r}_B- \mathbf{r}_0,i\xi)
\right].
\label{eq:U_atom_atom}
\end{equation}

\subparagraph{Non-retarded regime.}
When the interatomic separation satisfies $h\ll c/\omega_0$, retardation effects can be neglected. In this limit, the Green's tensor reduces to its quasi-static form,
\begin{equation}
\doubleoverline{\mathrm{G}}^{\mathrm{(0)}}(\mathbf{r},i\xi)
\propto
\frac{1}{r^3}.
\end{equation}
Substituting this expression into Eq.~\eqref{eq:U_atom_atom}, the interaction energy scales as
\begin{equation}
U_{\mathrm{CP}}(h)
\propto
-\int_0^\infty d\xi\;
\alpha(i\xi)^2\,
\frac{1}{h^6}.
\end{equation}
The frequency integral is controlled by the intrinsic spectral width of the atomic polarizability and yields a constant factor of order $\alpha(0)^2\omega_0$. Consequently,
\begin{equation}
U_{\mathrm{CP}}(h)\propto -\frac{1}{h^6},
\qquad
F_{\mathrm{CP}}(h)=-\partial_h U_{\mathrm{CP}}(h)\propto -\frac{1}{h^7}.
\end{equation}

\subparagraph{Retarded regime.}
In the retarded limit $h\ll \lambda_0$, the dominant contribution to the frequency integral arises from imaginary frequencies $\xi\sim c/h\ll \omega_0$, such that $\alpha(i\xi)\simeq \alpha(0)$. 
At imaginary frequencies, the dyadic Green's tensor exhibits the asymptotic scaling
\begin{equation}
\doubleoverline{\mathrm{G}}^{\mathrm{(0)}}(\mathbf{r},\omega)
\propto
\frac{e^{-\kappa r}}{r^3},
\qquad
\kappa=\frac{\xi}{c},
\end{equation}
which yields
\begin{equation}
\mathrm{Tr}
\!\left[
\doubleoverline{\mathrm{G}}^{\mathrm{(0)}}
\cdot
\doubleoverline{\mathrm{G}}^{\mathrm{(0)}}
\right]
\propto
\frac{e^{-2\kappa h}}{h^6}.
\end{equation}
Substituting into Eq.~\eqref{eq:U_atom_atom}, the interaction energy scales as
\begin{equation}
U_{\mathrm{CP}}(h)
\propto
-\alpha(0)^2
\int_0^\infty d\xi\;
\frac{e^{-2\kappa h}}{h^6}.
\end{equation}
Introducing the dimensionless variable $u=\xi h/c$, one finds
\begin{equation}
U_{\mathrm{CP}}(h)
\propto
-\frac{\alpha(0)^2 c}{h^7}
\int_0^\infty du\; e^{-2u}
\end{equation}
which leads to the retarded CP force
\begin{equation}
F_{\mathrm{CP}}(h)=-\partial_h U_{\mathrm{CP}}(h)\propto -\frac{1}{h^8}.
\end{equation}

These results recover the well-known crossover of the Casimir-Polder force from the non-retarded regime to the retarded regime between two atoms.

\subsection*{B. Atom--Array interaction}

We now extend the scattering approach to the interaction between a single atom and a 2D atomic array.
Within this formulation, the CP interaction admits a systematic expansion in powers of the product 
$\alpha(i\xi)\,\doubleoverline{\mathrm{G}}^{\mathrm{(0)}}$, corresponding to successive multiple-scattering processes of virtual photons.
Importantly, along the imaginary-frequency axis the atomic polarizability remains bounded and free of resonant enhancement, while the Green's tensor decays rapidly with both distance and frequency. As a result, higher-order scattering terms are strongly suppressed over a broad range of interatomic spacings. The interaction energy is therefore well captured by the second-order contribution, which represents a single round-trip scattering process. Upon neglecting higher-order terms, symmetry implies that the atom--array and array--array configurations exhibit identical distance scaling, so that the atom--array geometry provides a minimal and transparent setting for revealing the physical origin of the modified scaling behaviors. As before, we consider an isotropic atom `$0$' located at
\begin{equation}
\mathbf{r}_0=(0,0,h),
\end{equation}
interacting with a 2D infinite array $B'$ composed of identical atoms positioned at
\begin{equation}
\mathbf{r}_{mn}=(m a, n a, 0),
\end{equation}
where $a$ is the lattice constant and $m,n\in\mathbb{Z}$.
Within the scattering formalism, the fluctuating dipole moment of atom `$0$' satisfies
\begin{equation}
\label{eq:pA_array}
\mathbf{p}(\mathbf{r}_0)
=
\mathbf{p}_0(\mathbf{r}_0)
-
\frac{4\pi^2}{\epsilon_0\lambda^2}
\alpha(i\xi)
\sum_{m,n}
\doubleoverline{\mathrm{G}}^{\mathrm{(0)}}(\mathbf{r}_0-\mathbf{r}_{mn},\omega)
\cdot
\mathbf{p}(\mathbf{r}_{mn}),
\end{equation}
while the dipole moment of the atom at site $(m,n)$ in the array reads
\begin{equation}
\label{eq:pmn_array}
\mathbf{p}(\mathbf{r}_{mn})
=
\mathbf{p}_0(\mathbf{r}_{mn})
-
\frac{4\pi^2}{\epsilon_0\lambda^2}
\alpha(i\xi)
\doubleoverline{\mathrm{G}}^{\mathrm{(0)}}(\mathbf{r}_{mn}-\mathbf{r}_0,i\xi)
\cdot
\mathbf{p}(\mathbf{r}_0)
-
\frac{4\pi^2}{\epsilon_0\lambda^2}
\alpha(i\xi)
\!\!\sum_{(m'n')\neq(mn)}
\!\!
\doubleoverline{\mathrm{G}}^{\mathrm{(0)}}(\mathbf{r}_{mn}-\mathbf{r}_{m'n'},i\xi)
\cdot
\mathbf{p}_0(\mathbf{r}_{m'n'}) .
\end{equation}
To extract the leading-order CP interaction, we retain only the lowest-order scattering process in which vacuum fluctuations emitted by atom `$0$' are scattered once by the array and return back to atom `$0$'. Accordingly, we neglect multiple scattering events within the array and approximate
\begin{equation}
\mathbf{p}(\mathbf{r}_{mn})
\simeq
-\frac{4\pi^2}{\epsilon_0\lambda^2}
\alpha(i\xi)\,
\doubleoverline{\mathrm{G}}^{\mathrm{(0)}}(\mathbf{r}_{mn}-\mathbf{r}_0,i\xi)
\cdot
\mathbf{p}(\mathbf{r}_0).
\end{equation}
Substituting this expression into Eq.~\eqref{eq:pA_array}, the self-energy correction of atom `$0$' is governed by the kernel
\begin{equation}
\sum_{m,n}
\doubleoverline{\mathrm{G}}^{\mathrm{(0)}}(\mathbf{r}_0-\mathbf{r}_{mn},i\xi)
\cdot
\doubleoverline{\mathrm{G}}^{\mathrm{(0)}}(\mathbf{r}_{mn}-\mathbf{r}_0,i\xi),
\end{equation}
which is the natural generalization of the atom--atom result to a many-body environment.
Upon transforming to imaginary frequencies and averaging over vacuum fluctuations, the atom--array CP interaction energy can be written as
\begin{equation}
\label{eq:U_atom_array}
U_{\mathrm{CP}}(h)
\propto
-\int_0^\infty d\xi\;
\alpha(i\xi)\alpha(i\xi)
\sum_{m,n}
\mathrm{Tr}
\!\left[
\doubleoverline{\mathrm{G}}^{\mathrm{(0)}}(\mathbf{R}_{mn},i\xi)
\cdot
\doubleoverline{\mathrm{G}}^{\mathrm{(0)}}(\mathbf{R}_{mn},i\xi)
\right],
\end{equation}
where
\begin{equation}
\mathbf{R}_{mn}
=
\mathbf{r}_{mn}-\mathbf{r}_{\mathrm{0}}.
\end{equation}
\subparagraph{Non-retarded regime.}
In the non-retarded regime $h\ll \lambda_0$, retardation effects are negligible and the free-space Green's tensor assumes its quasi-static form
\begin{equation}
\doubleoverline{\mathrm{G}}^{\mathrm{(0)}}(\mathbf{r},i\xi)
\propto
\frac{1}{r^3},
\end{equation}
independent of frequency. Substituting this into Eq.~\eqref{eq:U_atom_array}, the interaction energy becomes
\begin{equation}
U_{\mathrm{CP}}(h)
\propto
-\int_0^\infty d\xi\;
\alpha(i\xi)^2
\sum_{m,n}
\frac{1}{r_{mn}^6},
\qquad
r_{mn}=\sqrt{h^2+(ma)^2+(na)^2}.
\end{equation}
To determine the distance scaling, we analyze the discrete lattice sum
\begin{equation}
S(h)=\sum_{m,n}\frac{1}{\bigl[h^2+(ma)^2+(na)^2\bigr]^3}.
\end{equation}
which can be rewritten as 
\begin{equation}
S(h)
=
\frac{1}{h^6}
\sum_{m,n}
\frac{1}{\bigl[1+(ma/h)^2+(na/h)^2\bigr]^3}.
\end{equation}
In the limit of $h \gg a$, the lattice spacing becomes much smaller than the
characteristic length scale set by $h$, and the discrete sum can be
approximated by a continuum integral.
Introducing dimensionless variables
\[
x = \frac{m a}{h} \qquad \mathrm{and}\qquad y = \frac{n a}{h},
\]

Replacing the summation by an integral according to
$
\sum_{m,n} \;\longrightarrow\; \frac{h^2}{a^2}\int \mathrm{d}x\,\mathrm{d}y,
$
we have
\begin{equation}
S(h)
\simeq
\frac{1}{h^6}\frac{h^2}{a^2}
\int_{-\infty}^{\infty}\!\!\mathrm{d}x
\int_{-\infty}^{\infty}\!\!\mathrm{d}y\,
\frac{1}{(1+x^2+y^2)^3}.
\end{equation}
The remaining integral is convergent and can be evaluated in polar
coordinates, yielding
\begin{equation}
\int_{-\infty}^{\infty}\!\!\mathrm{d}x
\int_{-\infty}^{\infty}\!\!\mathrm{d}y\,\frac{1}{(1+x^2+y^2)^3}
=
2\pi \int_0^\infty \frac{r\,\mathrm{d}r}{(1+r^2)^3}
=
\frac{\pi}{2}.
\end{equation}
Direct numerical evaluation of the discrete lattice sum shows that the
continuum approximation is already quantitatively accurate for separations
as small as $h \gtrsim a$, with deviations confined to an overall numerical
prefactor. Thus, the lattice sum can be expressed in
the asymptotic form
\begin{equation}
S(h)
\propto
\frac{1}{a^2\,h^4},
\qquad (h \gtrsim a),
\end{equation}
up to a prefactor of order unity.

In the limit $h \ll a$, the atom--array separation
is much smaller than the lattice spacing. In this regime, the discreteness of
the lattice becomes essential and the continuum approximation breaks down. $S(h)$ is then dominated by the nearest lattice site located directly beneath atom
$A$, i.e., $(m,n)=(0,0)$. All other lattice sites satisfy
$r_{mn}\gtrsim a\gg h$ and therefore give negligibly  smaller contributions.

Separating out the nearest-neighbor term, one finds
\begin{equation}
S(h)
=
\frac{1}{h^6}
+
\sum_{(m,n)\neq(0,0)}
\frac{1}{\bigl[(ma)^2+(na)^2\bigr]^3}
\left[1+O\!\left(\frac{h^2}{a^2}\right)\right].
\end{equation}
The second term converges to a finite constant independent of $h$, while the
first term diverges as $h^{-6}$ and dominates in the limit
$h\ll a$. Consequently, we have
\begin{equation}
U_{\mathrm{CP}}(h)\propto -S(h)\propto -\frac{1}{h^6} \quad \mathrm{and}\quad F_{\mathrm{CP}}(h)=-\partial_h U_{\mathrm{CP}}(h)\propto -\frac{1}{h^7},
\qquad \mathrm{for} \ h\ll a.
\end{equation}

However, when $h\gtrsim a$, as demonstrated from Eq. (S61) to (S63), contribution from the multi-site effect in the array becomes
remarkable and the collective nature of the array leads to a slower decay as
\begin{equation}
U_{\mathrm{CP}}(h)\propto -\frac{1}{a^2 h^4} 
\quad\mathrm{and}\quad
F_{\mathrm{CP}}(h)\propto -\frac{1}{a^2 h^5}.
\end{equation}
This analysis shows that even in the non-retarded regime, due to the discreteness of the atomic array, the CP
interaction exhibits a crossover from an atom--atom–like scaling to a
collective atom--array scaling as we see in the Rydberg case.

\subparagraph{Retarded regime.}

We now consider the retarded regime $h\gg c/\omega_0$, where the retardation effect
becomes essential. In this limit, the dominant contribution to the frequency
integral arises from imaginary frequencies $\xi\sim c/h\ll \omega_0$, such that
the atomic polarizability can be approximated by its static value,
$\alpha(i\xi)\simeq \alpha(0)$.

At imaginary frequencies, the free-space dyadic Green's tensor exhibits the
asymptotic form
\begin{equation}
\doubleoverline{\mathrm{G}}^{\mathrm{(0)}}(i\xi,\mathbf{r})
\propto
\frac{e^{-\kappa r}}{r^3},
\qquad
\kappa=\frac{\xi}{c},
\end{equation}
which yields
\begin{equation}
\mathrm{Tr}
\!\left[
\doubleoverline{\mathrm{G}}^{\mathrm{(0)}}
\cdot
\doubleoverline{\mathrm{G}}^{\mathrm{(0)}}
\right]
\propto
\frac{e^{-2\kappa r}}{r^6}.
\end{equation}
Substituting this expression into Eq.~\eqref{eq:U_atom_array}, the interaction
energy between atom `$0$' and the array reads
\begin{equation}
U_{\mathrm{CP}}(h)
\propto
-\alpha(0)^2
\int_0^\infty d\xi
\sum_{m,n}
\frac{e^{-2\kappa r_{mn}}}{r_{mn}^6},
\qquad
r_{mn}=\sqrt{h^2+(ma)^2+(na)^2}.
\end{equation}
In this case, $h\gtrsim a$ is always satisfied, so the summation over lattice sites can be approximated by a continuum integral in the transverse plane,
\begin{equation}
\sum_{m,n}
\;\longrightarrow\;
\frac{1}{a^2}\int d^2\boldsymbol{\rho},
\qquad
r=\sqrt{h^2+\rho^2}.
\end{equation}
Performing the angular integration and introducing the dimensionless variables
$u=\xi h/c$ and $x=\rho/h$, one finds
\begin{equation}
U_{\mathrm{CP}}(h)
\propto
-\frac{\alpha(0)^2}{a^2}
\int_0^\infty d\xi
\int_0^\infty d\rho\;
\frac{\rho\,e^{-2\kappa\sqrt{h^2+\rho^2}}}
{\left(h^2+\rho^2\right)^3}
=
-\frac{\alpha(0)^2 c}{a^2 h^5}
\int_0^\infty du
\int_0^\infty dx\;
\frac{x\,e^{-2u\sqrt{1+x^2}}}
{(1+x^2)^3}.
\end{equation}
The remaining double integral converges and can be evaluated as
\begin{equation}
\int_0^\infty du
\int_0^\infty dx\;
\frac{x\,e^{-2u\sqrt{1+x^2}}}
{(1+x^2)^3} = \frac{1}{10}.
\end{equation}
Then, this yields the the asymptotic scaling in the retarded regime as
\begin{equation}
U_{\mathrm{CP}}(h)\propto -\frac{1}{a^2 h^5},
\qquad
F_{\mathrm{CP}}(h)=-\partial_h U_{\mathrm{CP}}(h)\propto -\frac{1}{a^2 h^6}.
\end{equation}
\end{document}